\documentclass[12pt,preprint]{aastex}

\def\etal{et al.\rm}

\begin{document}
 
\title{Around the Clock Observations of the Q0957+561 A,B Gravitationally
Lensed Quasar II: Results for the second observing season}

\author{
Wesley N. Colley\altaffilmark{1},
Rudolph E. Schild\altaffilmark{2},
Cristina Abajas\altaffilmark{3},
David Alcalde\altaffilmark{3},
Zeki Aslan\altaffilmark{4,5},
Ilfan Bikmaev\altaffilmark{6},
Vahram Chavushyan\altaffilmark{7},
Luis Chinarro\altaffilmark {3},
Jean-Philippe Cournoyer\altaffilmark{8},
Richard Crowe\altaffilmark{9},
Vladimir Dudinov\altaffilmark{10},
Anna Kathinka\altaffilmark{11},
Dalland Evans\altaffilmark{11},
Young-Beom Jeon\altaffilmark{12},
Luis J. Goicoechea\altaffilmark{13}, 
Orhan Golbasi\altaffilmark{5},
Irek Khamitov\altaffilmark{5},
Kjetil Kjernsmo\altaffilmark{11},
Hyun Ju Lee\altaffilmark{12},
Jonghwan Lee\altaffilmark{12,14},
Ki Won Lee\altaffilmark{12,15},
Myung Gyoon Lee\altaffilmark{12},
Omar Lopez-Cruz\altaffilmark{7,16,17},
Evencio Mediavilla\altaffilmark{3},
Anthony F.J. Moffatt\altaffilmark{8, 18},
Raul Mujica\altaffilmark{7},
Aurora M. Ullan\altaffilmark{13},
Alexander Oscoz\altaffilmark{3},
Myeong-Gu Park\altaffilmark{19},
Norman Purves\altaffilmark{9},
Nail Sakhibullin\altaffilmark{6},
Igor Sinelnikov\altaffilmark{10},
Rolf Stabell\altaffilmark{11},
Alan Stockton\altaffilmark{9},
Jan Teuber\altaffilmark{20}, 
Roy Thompson\altaffilmark{9},
Hwa-Sung Woo\altaffilmark{19}, and
Alexander Zheleznyak\altaffilmark{10}}

\altaffiltext{1}{Dept. of Astronomy, University of Virginia, P.O. Box 3818,
Charlottesville, VA 22903}

\altaffiltext{2}{Harvard-Smithsonian Center for Astrophysics, 60 Garden St.,
MS-19, Cambridge, MA 02138}

\altaffiltext{3}{Instituto de Astrofisica de Canarias, Via Lactea, E-38200
La Laguna, Tenerife Canary Islands, Spain}

\altaffiltext{4}{Akdeniz University, Physics Department, 07058 Antalya, Turkey}

\altaffiltext{5}{Tubitak National Observatory, Akdeniz Universitesi Kampusu,
07058 Antalya, Turkey}


\altaffiltext{6}{Kazan State University, Russia}

\altaffiltext{7}{Insituto Nacional de Astrof\'{\i}sica Optica y Electr\'onica,
A.P. 51 y 216, C.P. 72000, Puebla, Pue., M\'exico}

\altaffiltext{8}{Department de Physique, Universite de Montreal, C.P. 6128,
Succ. Centre-Ville, Montreal, QC, H3C 3J7}

\altaffiltext{9}{Institute for Astronomy, University of Hawaii}

\altaffiltext{10}{Kharkov National University, Astronomical Observatory,
Sumska Str. 35, Kharkov, 61022, Ukrania}

\altaffiltext{11}{University of Oslo, Institute of Theoretical
Astrophysics, PO Box 1029, Blindern, Oslo, N-0315, Norway} 

\altaffiltext{12}{Seoul National University, Astronomy Program, SEES, Seoul,
151-742, Korea} 

\altaffiltext{13}{Departamento de Fisica Moderna, Universidad de Cantabria,
E-39005 Santander, Cantabria, Spain}

\altaffiltext{14}{DACOM Corporation, Internet Technology Division, DACOM
Bldg., 706-1, Yeoksam-dong, Kangnam-ku, Seoul 135-610, Rep. of Korea}

\altaffiltext{15}{University College London, UK}

\altaffiltext{16}{Visiting Researcher, Departamento de
Astronom{\'\i}a, Universidad de Gunanajuato, M\'exico}

\altaffiltext{17}{Fulbright Visiting Scholar at the Observatories of
the Carnegie Institution of Washington (OCIW), U.S.A.}

\altaffiltext{18}{Observatoire du mont Megantic, Canada}

\altaffiltext{19}{Kyungpook National University, Korea}

\altaffiltext{20}{Danish Library for Natural and Medical Sciences, Copenhagen,
Denmark}






 
\begin{abstract}

We report on an observing campaign in March 2001 to monitor the brightness of
the later arriving Q0957+561 B image in order to compare with the previously
published brightness observations of the (first arriving) A image. The 12
participating observatories provided 3543 image frames which we have analyzed
for brightness fluctuations. From our classical methods for time delay
determination, we find a $417.09 \pm 0.07$ day time delay which should be free
of effects due to incomplete sampling.  During the campaign period, the quasar
brightness was relatively constant and only small fluctuations were found; we
compare the structure function for the new data with structure function
estimates for the 1995--6 epoch, and show that the structure function is
statistically non-stationary.  We also examine the data for any evidence of
correlated fluctuations at zero lag.  We discuss the limits to our ability to
measure the cosmological time delay if the quasar's emitting surface is time
resolved, as seems likely.

\end{abstract}
\section{Introduction}\label{sec:Intr}

An observing campaign to determine the Q0957+561 A,B gravitational lens
time delay to a fraction of a day has been undertaken, justified by the
evidence available that the quasar has brightness fluctuations on time
scales of hours and that microlensing on day timescales is observed (Colley
\& Schild 1999). Our report (Colley \etal\ 2002) of the brightness
fluctuations observed in Jan 2000 in the first arriving A image becomes a
prediction of the pattern of fluctuations expected in March 2001. In this
report we present reductions of CCD images for the determination of the B
image brightness record, and the determination of a refined value of the
time delay.

This time delay determination comes as refinement of previous estimates which
have gradually converged to a value near 417 days.  After some years of
uncertainty whether the delay was near 415 days or 540 days, new estimates have
come from new data sets and re-analysis of extensive older data sets to produce
values of 416.3 (Pelt \etal\ 1998), 417.5 (Kundi\'c \etal\ 1997), $425 \pm 17$
(Pijpers 1997), and 417.4 (Colley \& Schild 2000).  Re-analysis of much of
the same data has produced a divergent value of $422.6 \pm 0.6$ days (Oscoz et
al, 1999) but our present program allows no check on this value because our
monitoring was only over a 10 day time interval.  Radio brightness monitoring
has not had the time sampling, accuracy, or the demonstrated rapid variability
to justify participation in this program.

In section 2 we report the new data and reductions for the 12 participating
observatories. Section 3 contains an analysis for time delay. We found only a
low-amplitude of brightness fluctuations for the quasar during the monitoring
period, and any evidence for rapid microlensing is within the noise of our
data. However a sharpened value of the time delay allows us to re-analyze high
quality data sets from published reports (Colley \& Schild, 2000), and a
determination of the structure function for the rapid microlensing in that
dataset will be the subject of a subsequent report.

\section{Observations}

Our list of participating observatories was shown as Table 1 of Colley
\etal\ (2002), (Paper 1). For the second season three additional
observatories joined the collaboration, principally to cover the Pacific
region. The 61cm Mauna Kea reflector was operated from the Institute for
Astronomy offices at Hilo, Hawaii.  The Mexican 1.5m Harold Johnson
Telescope at OAN (Observatorio Astron\'omico Nacional) at San Pedro
M\'artir joined the 2.1m OAGH telescope in northern Mexico. The
Mt.~Megantic telescope in Montreal, Canada provided coverage of the North
American continent from a different weather zone.

In this section we provide additional remarks about the data reductions and
its relationship to the telescope and camera properties. Any changes from
the properties noted in Paper 1 will be given. Our list begins at the
international date line and is ordered by increasing longitude. Throughout
this section and report, we refer to R filter images only, and although
approximately 10\% of our data was taken with a V filter we do not remark
on the results for this supplementary reduced data set.

\subsection{BOAO (S. Korea)} 

This observatory produced 180 quasar image frames over 5 nights, all of
uniformly high quality. The BOAO data from our first year observations are
particularly important because they recorded a rapid brightness decline in
the A image that will strongly contribute to the present time delay
determination. However, as noted in Paper 1, the rapid brightness decline
seemed to be recorded in both quasar images and we were suspicious that it
might have been the result of some peculiar instrumental effect, although
re-reduction of the data with entirely different software seemed to confirm
the reality of the rapid brightness change. Thus it is reassuring that the
BOAO data in the new campaign seem to again be of high quality and contain
no bothersome artifacts evident in the reduced data.

\subsection{SOAO (S. Korea)}

This 61cm telescope on the top of Mt.~Sobak in the middle of South Korea is
often differently affected by weather than BOAO, and so is of importance to us
because of concern about coverage of the vast Pacific region. The telescope
lacks an offset guider, however, and some of the images are streaked
slightly. Nevertheless we reduced 236 quasar image frames obtained over 5
nights, albeit with slightly shorter exposure times. The smaller aperture
produces lower signal levels, and the error bars for the data set may be seen
to be somewhat larger.

However the time coverage by the SOAO observatory has been critical to our
program, especially as it provided the only coincident coverage of the rapid
brightness decline recorded the previous year at BOAO as noted previously.
The relatively large amplitude, approximately 40 milli-magnitudes (mmag)
over just 5 hours, adds considerable weight to our time delay
determination. 

\subsection{Maidanak (Uzbekistan)}

Since the previous year's campaign, the Maidanak 1.5m telescope has been
equipped with a new CCD camera having $2048\times 800$ pixels and providing an
image scale of .121 arcsec/pixel. The images are of the highest quality because
of a combination of superb optics and consistently excellent seeing.

We did, however, encounter a problem with the new camera. A large diffuse
bright spot of only 3\% amplitude covers the entire central region of the
CCD frame. The spot is apparently caused by light that is diffusely
reflected from the region of the CCD detector back to the camera optics,
and then again reflected back onto the CCD camera. This produces an excess
brightness near the center of the CCD frame that the standard
flat-fielding procedure interprets as locally more sensitive pixels.
So in correcting for the apparently high sensitivity of the central
region's pixels, the flat-fielding procedure also reduces the brightnesses
of the central stars in the corrected frame. The point is that
flat-fielding assumes that all processes affecting the sky brightness
across the image frame, such as pixel-to-pixel sensitivity and vignetting,
are multiplicative, and a process that adds brightness causes failures
and errors in the flat-fielding process. We corrected the problem by
preparing an unsharp mask image from a stack of nighttime image frames, and
subtracting off the excess brightness from our image frames and flat-field
frames before flat-fielding.

The corrected frames have produced an important backbone to our brightness
record, because of the consistently high quality and because of the
excellent coverage of our monitoring window. The campaign produced 395
image frames over 8 nights in our program.

\subsection{TUG (Turkey)} 

The 1.5m telescope was not equipped with an offset guider, and some images
are somewhat streaked but the effect was minimized by taking relatively
short exposures. A total of 192 image frames were collected on 6 nights.
We were unable to obtain consistent photometry from the images, and we do
not believe that the guiding was the problem. Instead, it appears that a
slight non-linearity in the camera's response is indicated. For example, on
the cloudiest night, although the quasar brightness is referenced to local
field stars, our reduced photometry was several percent different from
the photometry on strictly clear nights.

We are hopeful that the data will be reducible after some mapping function
of the camera response to light is available.  For the present cannot
include this otherwise excellent data set in our time delay determination or
in the Fig.~1 plot of the brightness curve, at least until we understand
better the camera's slightly non-linear response to light.

\subsection{NOT (Canary Islands)}

The 2.5m NOT telescope is the largest involved in our campaign and was
scheduled for 4 nights which turned out to have nearly ideal weather,
producing a harvest of 755 image frames. These have provided excellent
photometry and excellent statistics on the four nights, and together with
the Maidanak results have produced a ``backbone'' against which we refer
other observatories to look for systematic differences.

\subsection{JKT (Canary Islands)}

The 1m Johannes Kepler Telescope was scheduled for the last two nights of
the formal monitoring period plus two nights beyond to allow some
information about the time delay in case the 424-day delay value championed
by Oscoz \etal\ (2001) turn out to be correct. A total of 147 quasar image
frames were analyzed and produced a brightness record compatible with
results for the other observatories.

\subsection{IAC80 (Canary Islands)}

The 80cm reflector produced 173 quasar image frames on 8 nights of the 
formal monitoring period and 3 nights beyond to allow further check on the
longer delay value of 424 days advocated by Oscoz \etal\ (2001). We have not
been able to bring the photometric results into agreement with the
reductions for other observatories for reasons that we do not yet
understand. Most problematic are the results for JD2451983, where the
reduced photometry is 2\% brighter for image B than for image A as compared
to results for the same night from Maidanak and from NOT. Data from the
remaining nights seem not to share this defect, and we are puzzled about
its origin. Because the images were precisely registered on the CCD
detector by an offset guider, it is possible that a CCD defect has affected
the photometry for a single night. We have found that pixel-to-pixel
sensitivity variations of the CCD detector of 10\% are routinely found, and
we suspect that the CCD camera is not operated in correct adjustment.
We expect to investigate this anomaly
further, but for the present analysis we cannot justify censoring a single
night's data only, and for now we have not included the IAC data in
our final data compilation and time delay calculation.

\subsection{Mt.~Megantic (Montreal)}

The Mt.~Megantic observatory is sited at a very dark mountainous region 250 km
east of Montreal, near the U.S. border.  It offers a 1.6m Ritchey-Chretien
telescope with an offset guider that has been extensively used in photometric
programs to date. Over 4 nights the observatory contributed 115 data images of
excellent quality to our program.

\subsection{Mt.~Hopkins (Arizona)}

The Mt.~Hopkins Observatory has been the mainstay of Q0957 brightness
monitoring for over 20 years. Because of scheduling difficulties the 14
nights allocated overlap only 3 nights with the campaign time interval.
The additional 11 nights precede the campaign and allow check on possible
time delays less than the nominal 417 days. Over 12 nights 375 image
frames were reduced for photometry.

\subsection{Mexico - 1.5m Harold Johnson Telescope, San Pedro M\'artir (Mexico
2)} 

A total of 126 images over 4 nights were reduced for photometry with this
telescope. The images were of excellent quality. The telescope was
scheduled for only the first 7 nights of the campaign, with the last 3
nights covered by a second Mexican telescope.

\subsection{Mexico - 2.1m Telescope OAGH, Cananea (Mexico 1)}

From the OAGH (Observatorio Astrof{\'\i}sico Guillermo Haro) in Cananea,
Sonora, a total of 84 image frames were obtained over 2 nights at the end of
the monitor period.  Very poor (3.5 arcsec) seeing was experienced during part
of one night.  In general the image quality was not quite as good as the
upgraded Harold Johnson 1.5m telescope, likely a result of a problem with the
mercury belt of the mirror support system.  Nonetheless, the data reduction
procedure seems to have produced excellent results for this telescope.

\subsection{61cm Mauna Kea Telescope (Hawaii)}

The Hawaiian 61cm telescope was operated remotely from the Institute for
Astronomy at Hilo and from the University of Hawaii at Honolulu. With clear
skies and excellent seeing, a large quantity of data was obtained, but lack
of an offset guider caused some trailing of the images. Nevertheless most
of the data could be easily reduced with our robust computer program, and
the Hawaii telescope produced one of our main data sets. Over 8 nights 430
useful image frames were obtained.

\section{Analysis: Data Reduction and Time Delay}

The data for all observatories was analyzed by a single program as described in
Colley \etal\ (2002a,b) and in Colley \& Schild (1999, 2000).  Briefly, all
images were de-biased and flat fielded, and the corrected images had the star
positions identified from an automatic procedure. Aperture photometry was
performed on the two quasar images and several nearby standards, and the quasar
brightness was referenced to the standards.  Corrections for the aperture
crosstalk were determined according to the Colley \etal\ (2002b) method: a
simple parabolic fit to the magnitude relative to the mean magnitude of the
run, vs.~the log of FWHM seeing is made, and that signal is subtracted out.
Colley \& Schild (1999) showed that a detailed correction involving galaxy
subtraction from HST data (Bernstein \etal\ 1997) and detailed A-B aperture
cross-talk corrections yielded a relation to seeing well described by this very
simple model.

The reduced data are shown in Fig.~1, where we plot the photometry obtained
in the campaign time frame 12--21 March 2001. Upper and lower plots show the
A and B image photometry, and the bottom panel shows as bars the time
coverage of each participating observatory, and at the bottom the total
coverage. The plotted data points show the hourly photometry means obtained
by each observatory, and an error bar calculated from the photon statistics
relevant to the detection (not a standard error relative to the mean). 

We are frankly disappointed by the low level of brightness fluctuations shown
by the late-arriving B image, which will now be compared with the A image data
for the previous year. Note that from simple inspection it may be seen that the
amplitude of fluctuations in image B is approximately half the level seen in
image A for 2001. Similarly, data for image A in 2000 show fluctuations only
half as large as those for image B.  Both the 2000A and 2001B patterns exhibit
less than half the amplitude of the pattern found for fluctuations in 1994--96
by Colley and Schild (2000). Thus the quasar has given us an opportunity to
demonstrate that accurate photometry and detection of a very low level
amplitude of brightness fluctuations could be produced with our methodology,
but we would have preferred stronger fluctuations.

The time delay calculation was undertaken with the ``PRH'' method (Press \etal\
1991), which, despite some controversy has become a standard utility.  The
method is based on the notion that the quasar variations exhibit a power-law
``structure-function,'' which is to say the expected magnitude variance of one
point from another on the light-curve is power-law in the time separation of
the points, (i.e. $V \propto |t_1 - t_2|^\alpha$).  This method allows one to
address the non-uniform time sampling of the data without direct interpolation.
From there, usual second-order Gaussian statistical methods are used to
construct a ``$\chi^2$'' statistic that reduces to the usual $\chi^2$ method if
there were no interpolation necessary.

The method is known to have problems, particularly if the data records are
affected by microlensing (a highly non-Gaussian signal) (Press \& Rybicki 1997,
Thomson \& Schild 1997, Schild 1996).  The method also has a propensity to
favor lags where there is the least data overlap (Colley \& Schild, 2000), but
because our data have an irregular sampling history, this is not expected to be
a problem.

Results for the PRH method test are shown in Fig.~2, where we plot $\chi^2$
as a function of lag between the A and B images. A small valley for 417.1
days is presumed to indicate the best value of time delay.  Toward the left of
the plot, the $\chi^2$ value declines chiefly because the main feature of the
image A light-curve ceases to overlap with the image B's shifted dates.

The agreement of the two quasar brightness curves for a 417.1-day lag is
shown in Fig.~3, where we plot as unfilled circles the image A data from
2000 and as filled circles the image B data from 2001. Error bars are
computed as described previously, and are calculated from the fundamental
photon statistics, not from the purely empirical departure of individual
points from the hourly means. Also shown is the ``error snake'' that shows
the width of the 1-$\sigma$ error interval computed by the PRH method as
part of the interpolation scheme. Note that the mean width of this snake is
only 5 milli-mag (henceforth mmag, where 1 mmag = .001 mag). A quick glance
shows that the true errors seem to be very close to the computed errors, in
the sense that more than half the hourly average points lie within the
1-$\sigma$ ``snake''.

The formal time delay value calculated for the project is $417.09\; \pm\;
0.07$ days. Inspection of Fig.~3 shows that a weak pattern of fluctuations is
seen throughout the campaign period, and a single fluctuation at $\mbox{JD}
-2449000$ = 2564.5 of 30 mmag amplitude predominates. We suspect that the
overall pattern and the single large event contribute about equally to the time
delay value. As was noted by Colley \etal\ 2002a, the event was seen in the
Korean/BOAO data from the first season and not entirely believed because the
data seemed to show a simultaneous event in the B data for 2000 also. However
re-reduction of the data with a different analysis program (IRAF) seemed to
show that the feature is real, and its repetition in 2001 make the case more
convincing.

With the data in Fig.~3 plotted for the best fit lag, the plot also becomes
a record of microlensing, in the sense that any significant differences
between the two brightness records indicates a pattern of fluctuations not
intrinsic to the quasar, and presumably originating in the lens galaxy. We
do not find that the Fig.~3 comparison makes a convincing case that any
microlensing has been detected at the 5 mmag level.  There is an appearance
of a peak for $\mbox{JD} -2449000 = 2564$, and we note that evidence for
this peak comes from two observatories (Maidanak and BOAO.). We have
seen evidence for a peak of similar amplitude and duration in the Q0957 B
data record for JD 2449704 as illustrated in Fig.~3 of Colley \& Schild
(2000). The detection of an event in our microlensing record is only based
upon 3 hourly average data points, each having approximately $2 \sigma$
significance, and we feel obliged to err on the side of conservatism and
claim no significant detection.

We do not consider that this proves that rapid microlensing does not exist;
our sharpened time delay of 417.1 days will allow us to show from previous
data records that data from a single observatory can overlap, and produce
microlensing information. This will be the subject of a further report.

\section{The Structure Function}

In Fig.~4 we show the structure function for the quasar's brightness
fluctuations during Jan 2000 for image A and March 2001 for image B.  In
Figure 4 the variance plotted as a function of lag is a squared quantity,
so the actual brightness fluctuations are the square root of the plotted
numbers. Thus for a lag of one day, either image component showed variance
of approximately $10^{-5}$, or the mean brightness fluctuation was $0.3
\times 10^{-2}$, or 3 mmag. So on average, the quasar brightness changed by
only 3 mmag during any 24-hour time interval.

This level of brightness fluctuation is extraordinarily low for this
quasar, as may be seen from comparison with the solid line which shows the
fit to the variance measured in 1995 (Colley \& Schild 2000, Fig.~5).  Thus
we were extremely unlucky that the date chosen for the beginning of our
monitoring for reasons of optimum observability turned out to coincide with
a period of low quasar activity. This allowed us to demonstrate the ability
to reduce data from multiple observatories and measure brightness with high
precision, but we would have preferred to find large fluctuations to firmly
establish a time delay and microlensing.

Fig.~4 quantitatively shows that the statistics of the quasar's 
brightness fluctuations are highly non-stationary. On timescales of 1--10
days the fluctuations were smaller than measured in 1995; on time scales of
a year, the fluctuations were actually twice as large as measured in 1995
(the data point for year lag is far off scale and not seen on this plot).

\section{The Correlation for Zero Lag}

In the course of monitoring Q0957, many groups have noticed a ``zero lag''
correlation between the A and B images (e.g., Kundi\'c \etal\ 1995).  This
``zero lag'' correlation is impossible by all models of the gravitational
lensing, and would require some kind of precisely aligned gravitational
wave in the Halo of our Galaxy or perhaps a cosmic string.  It is presently
interpreted as a frame-by-frame correlated error in the photometry.

The Kundi\'c \etal\ (1995) group endeavored with fair diligence to uncover
the source of such an error, examining moon phase, zenith angle, and many
other observational states for some correlation with the apparent
photometric errors, but encountered little success.  We have also noticed
that our data seem, upon inspection, to exhibit a similar ``zero lag''
correlation, even after our parametric correction for seeing, discussed
previously.  In particular, the PRH ``snake,'' a completely objective
interpolation, yields light-curves in which the coincidence of the many
shallow peaks and valleys is immediately striking to the eye.  We therefore
engaged ourselves in a great amount of effort to remove errors that might
be correlated with flux, sky background, location on CCD, and time of
night, but those efforts have yielded little improvement for most
observatories.

We show the effect quantitatively in Fig.~5 as the PRH $\chi^2$ (lower
$\chi^2$ means higher correlation) for lags in the vicinity of zero lag in
the two data sets; the solid curve shows the correlation for the 2001 data,
and the dashed curve shows the correlation for 2000.  Any correlations
would be expected to be quite random in the vicinity of zero lag if the
photometry were perfect.  However, in both cases, there is a $\chi^2$
trough near zero.  For 2001, the overall minimum is at precisely zero lag,
quite suggestive of a photometric problem.  For 2000, the case is more
muddled.  The trough nearest zero is closer to 0.1 days and is only the
sixth lowest trough, while the overall minimum is at around a one day lag.
While in the 2001 dataset, there seems to be compelling evidence of a
photometric problem, nothing compelling arises in the 2000 dataset, despite
identical photometric reduction methods.

Most likely, there is a similar photometric error correlation in both 2000 and
2001, but the true correlation of the light-curves adds slight destructive and
constructive interference to that signal (respectively).  We interpret Fig.~5,
therefore, as showing a zero lag correlation that is evidence of an as yet
uncorrected photometric problem at the few mmag level, residing on top of the
true correlation (PRH $\chi^2$) of the A and B light-curves.


\section{The Possibility of Multiple Time Delays}

Our brightness monitoring has produced a time delay of $417.09 \pm 0.07$ days,
and scant evidence for microlensing. During our monitoring campaign the
quasar was experiencing below-normal brightness variability, and very
possibly the low level of microlensing fluctuations measured is related.

At the time this project was organized, time delays for the gravitational
lens system seemed to be converging to a value near 417 days. Analysis of
the 17-year brightness history by Pelt \etal\ (1998) produced a delay of
$416.3 \pm 1.7$ days, and the Kundic \etal\ (1997) observation of a large
event seemed to make their 417.4 day delay unquestionable, although in
hindsight the quoted value was for {\it g}-filter data, and their {\it
r}-filter data gave 420.3 days. Finally the extensive monitoring over 7
consecutive nights reported in Colley \& Schild (1999) seemed to make a
convincing case that significant fluctuations were observed, and repeated
after a time delay of 417.4 days. Their Fig.~7 appeared to show ample
repeated fluctuations to define adequately the time delay.

At the same time, suspicions of a somewhat longer time delay have arisen.
The Pijpers time delay determination using a long extensive data base gave
$425 \pm 17$ days, but the error seemed to include the favored shorter
value.  But then a report by Oscoz \etal\ (2001) seemed to show a longer
delay for the same data sets, utilizing different statistical methods. A
new thesis by Ovaldsen (2002) with re-reduction of the original data frames
seems to show not only stronger evidence for the 424-day delay, but even a
small local anti-correlation bump in the delay curve where the favored
417-day lag should be. Thus we find it perplexing and frustrating that with
so many nights of overlapping data, it is difficult to find an enduring
time delay.

As noted by Pelt \etal\ 1996, the fine structure filtered out of the
brightness record does not give a time delay value (Pelt \etal\ 1996,
Fig.~10, 11), even though hundreds of nights of data overlap for any test
value of lag near 420 days (Pelt \etal\ 1996, Fig.~1). If microlensing were
not affecting the brightness records, it should be easy to determine the
time delay to a fraction of a day.

We believe that it is now appropriate to re-assess our assumptions and
methodology to ask what we are seeking and why we have such problems
finding it. We seek the time delay between the propagation paths for the
two quasar images, where it has long been believed that the geometry of the
smooth mass distribution in the direction of the lens system provides
different paths and the two quasar images are seen at different times.  We
have presumed that by observing brightness fluctuations of the two images
we will be able to match up the patterns and measure the time delay.  But
what are the limits to this procedure if the quasar's luminous structure is
so large that microlensing by massive objects in the lens galaxy amplifies
different parts of the quasar by different amounts?  What if, for example,
the inner edge of the accretion disc is highly microlensed for the A image
but not for B?

For the Q0957 radio source quasar the black hole mass would be perhaps $3
\times 10^9 M_{\odot}$, giving a Schwarzschild diameter of $2 \times
10^{15}$ cm. Thus the diameter of the innermost stable orbit, $6R_S$, is
approximately $10^{16}$ cm, or 4 light days. For a quasar redshift of 1.4,
a proper time of 4 light days is observed as 10 days.  As such, a maximum
of 1\% variability might be expected on the timescale of hours (pertinent
to this paper).  Colley \& Schild (1999) showed by measuring the structure
function down to sub-day lags, that more typically the quasar varies about
1\% per day on average.

With an inner diameter for the accretion disc of order 10 days (our time),
a simultaneous event that illuminated the entire inner disc would appear to
us to brighten roughly 10 days earlier on the front size compared to the
back side.  This simple light travel time couples strongly to the
microlensing.  By coincidence, the Einstein radius (at the source), is also
of order $10^{16}\;\mbox{cm}$ for stellar mass objects in the lens galaxy.
In this high optical depth case, the Einstein radius becomes the typical
separation between caustics.  Thus, if the stars alone had an optical depth
of order 0.1, of order 10\% of the time one would expect a caustic to be
lying somewhere on the inner accretion disc.  Hence, one image could
significantly magnify the same event at a different time than the other,
due to simple light travel time from one edge of the QSO emitting surface
to the other.

Since the inner edge of the disc presumably demonstrates a significant
fraction of the variability seen on the few-week timescale (typical of the
large events seen in the QSO historically, including those leveraged for
time-delay measurements [e.g., Kundi\'c \etal\ 1997]), a light travel
time problem could be significant even without a MACHO dark matter
population.  If there is a significant population of MACHOs, the problem
becomes nearly unavoidable.

Additional outer structure at scales of 100 days (our time) is implied in
the Elvis (1999) model.  If such structure responds to an event near the
center, the light travel time problems are exacerbated, particularly for
stellar mass lenses, since one would expect multiple caustics to lie
somewhere on this outer structure all the time.  Thomson \& Schild (1997)
encountered autocorrelation peaks near 100 proper light days, which
supports the Elvis (1999) model.  Such a model would necessarily present
substantial light travel time signal in the light curve.

Further evidence for the phenomenon comes from a cross-correlation
calculation for Q0957 by Schild and Cholfin (1986), who showed a Full Width
at Half Maximum (FWHM) of nearly 100 days; subsequent analyses of the same
or comparable data sets by Vanderriest \etal\ (1989) and Press, Rybicki,
and Hewitt (1991) gave a similar result.  Even the modern calculations by
Pelt \etal\ (1997) shows in their Fig.~3 and following figures a 100-day
wide correlation curve. Only the more modern calculations for short data
sets centered around strong brightness changes give the sharper
cross-correlation peaks of programs by Kundi\'c \etal\ (1997) and Colley \&
Schild (1999).  Thus, quasar structure on observed scales of 100 light-days
(observed) has long been predicted by observations.

More significant results were gleaned from a statistical analysis of the
brightness records by Thomson \& Schild (1997) who found two unexpected
facts. Their Fig.~6 shows that different subsets of the brightness data
show different lags, with the lags seeming to persist over approximately 2
years. This would be expected if microlensing were locally magnifying
different parts of the quasar's luminous region by different amounts as the
pattern of microlenses changes on a time scale of 2 years.  Furthermore,
their Fig.~3 seems to show autocorrelation peaks with observed lags near
200 days, suggesting that the quasar has some structure on scales much
larger than the 1-night sampling of most data. (note that for a quasar at
$z=1.4$, the proper scale of the implied quasar structure is 200 days /
2.4, or approximately 70 light day proper size scale.) Weaker structure on
smaller scales is further implied by autocorrelation peaks near 20 light
days (observed).

Thus we interpret our new time delay measurement as follows.  For the
microlensing configuration observed in calendar 2000--01, the time delay
corresponding to the microlensing alignment then relevant is 417.1
days. Possibly other parts of the quasar were being magnified to produce
other lags as well, but their signature is hardly apparent because of the
disappointingly low level of quasar activity.  Very probably a component of
what is commonly called rapid microlensing is actually the result of quasar
brightness fluctuations seen magnified by the different microlensing of the
A and B images, and any other data sets taken at epochs close to ours might
well find evidence for other lags. Our 417.1-day lag is sharply defined by
the depth and narrowness of the valley of our PRH autocorrelation
calculation, but it does not allow prediction of lags that will be observed
several years in the future. Use of this lag will allow us to make careful
comparisons of older data sets and allow some conclusions to be made about
the structure function for rapid microlensing, which will relate to both
the quasar's structure and the microlensing population.

\section{Conclusions and Discussion}

We have carried out the first gravitational lens time delay measurement
from data sampled nearly continuously over many days.  The measurement of
$417.09 \pm 0.07$ is certainly consistent with many previous efforts (e.g.,
Kundi\'c \etal\ 1997), but puzzlingly inconsistent with others (e.g., Oscoz
\etal\ 2001).  Since there was surprisingly little variation in the
brightness of the QSO images, compared to previous structure function
measurements (Colley \& Schild 2000), the time delay measurement is not as
strong as we had hoped.  Furthermore, there is little evidence of
microlensing on timescales of hours to days.

In recognizing that there are now (again) two credible, and competing time
delays (around 417 days and around 424 days), both with a wealth of
evidence to support them, one must attribute the discrepancy either to a
lack of quality data, as was done a decade ago (PRH), or to the possibility
that the QSO and lens system form multiple delays which foil our efforts to
produce a unique time delay.  The former seems increasingly untenable given
the level of observational effort poured into this system over the past
decade.  If the multiple delay hypothesis is to be examined, careful
consideration of more complex QSO models must be undertaken (e.g., Wyithe
\& Loeb 2002).

As a final comment, we note that one of the most intriguing aspects of this
work has been the formation of a large, international collaboration, and
integration of data from widely varying nations, telescopes and instruments.
As the number of lensed QSOs has grown, so has the need for nearly constant
monitoring by as many telescopes in as many locations as possible, with some
coordination.  Our project has demonstrated proof that a large international
collaboration of medium-sized observatories can be coordinated sufficiently to
obtain nearly constant monitoring of lensed QSOs.

\begin{acknowledgements}

WNC thanks the Smithsonian Astrophysical Observatory which provided the
computing resources for this project.  We wish to thank University of
Hawaii graduate students Kevin Sweeney, Jennifer Halsted, and Masa
Matulonis for assisting with the observing. Ki-Won Lee assisted in the SOAO
observations. Andrew Pickles was instrumental is setting up the Hawaii
telescope for remote observing.  MGP was supported by the Korea Science \&
Engineering Foundation, grant No. R01-1999-00023.  MGL was supported in
part by the Korea Research Foundation, Grant No. KRF-2000-DP0450.  AFJM and
J-PC are grateful for financial aid from NSERC (Canada) and FCAR (Quebec)
We wish to thank the OAN-SPM and specially the TAC president Margarita
Rosado for making the Mexican participation in this campaign possible. The
research of OLC, RM and VCH is partially supported by CONACYT grants
J32098-E, J32178-E and 28499-E, respectively.

\end{acknowledgements}

\begin{figure}[t]
\plotone{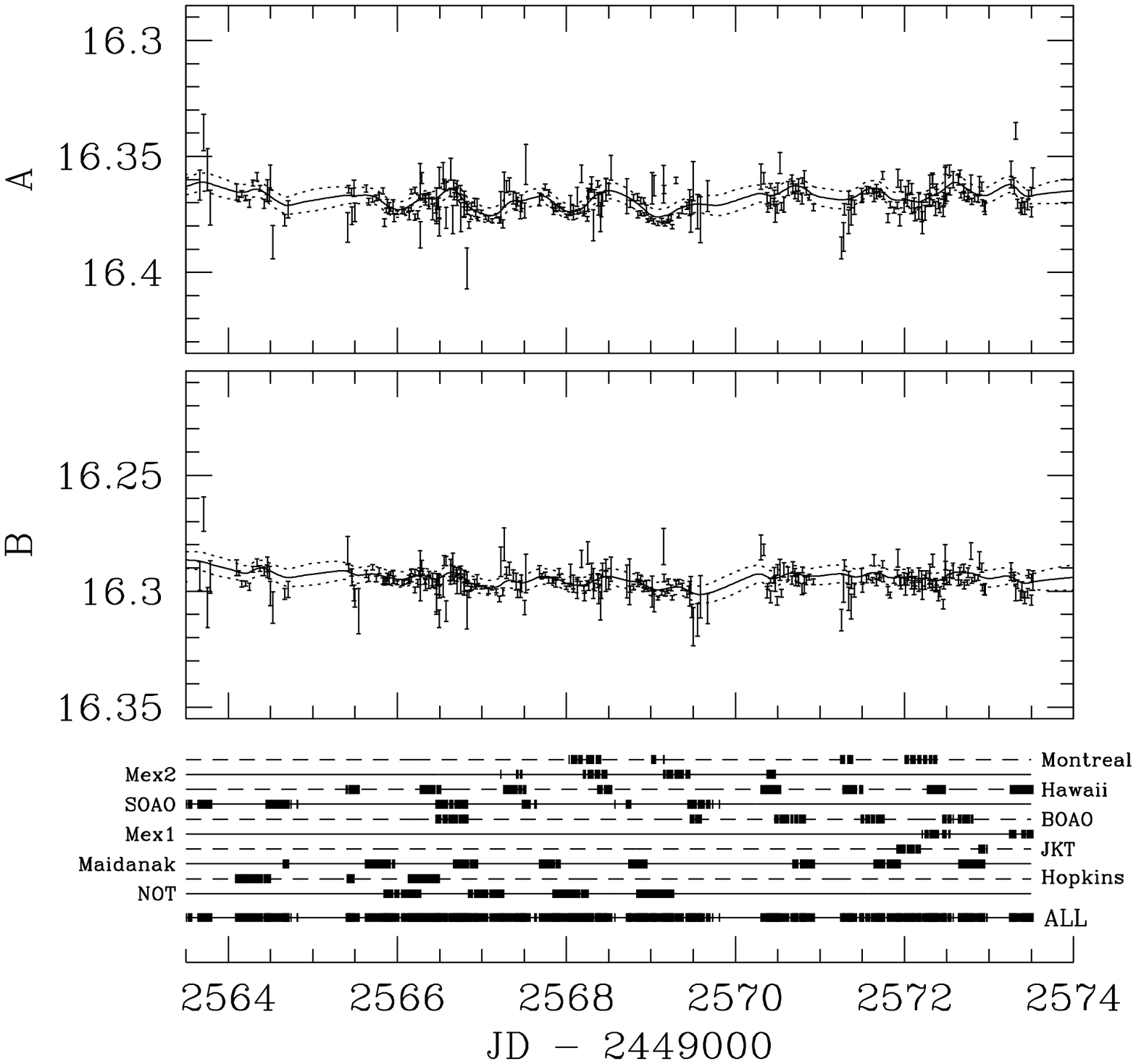}
\caption{The R-band light curve of Q0957+561A,B from March 12 to March 21,
2001.  At top is the image A brightness record; in the middle is the image B
brightness record, and at bottom is a series of line density graphs
illustrating when each observatory was contributing data.}
\label{fig1}
\end{figure}

\begin{figure}[t]
\plotone{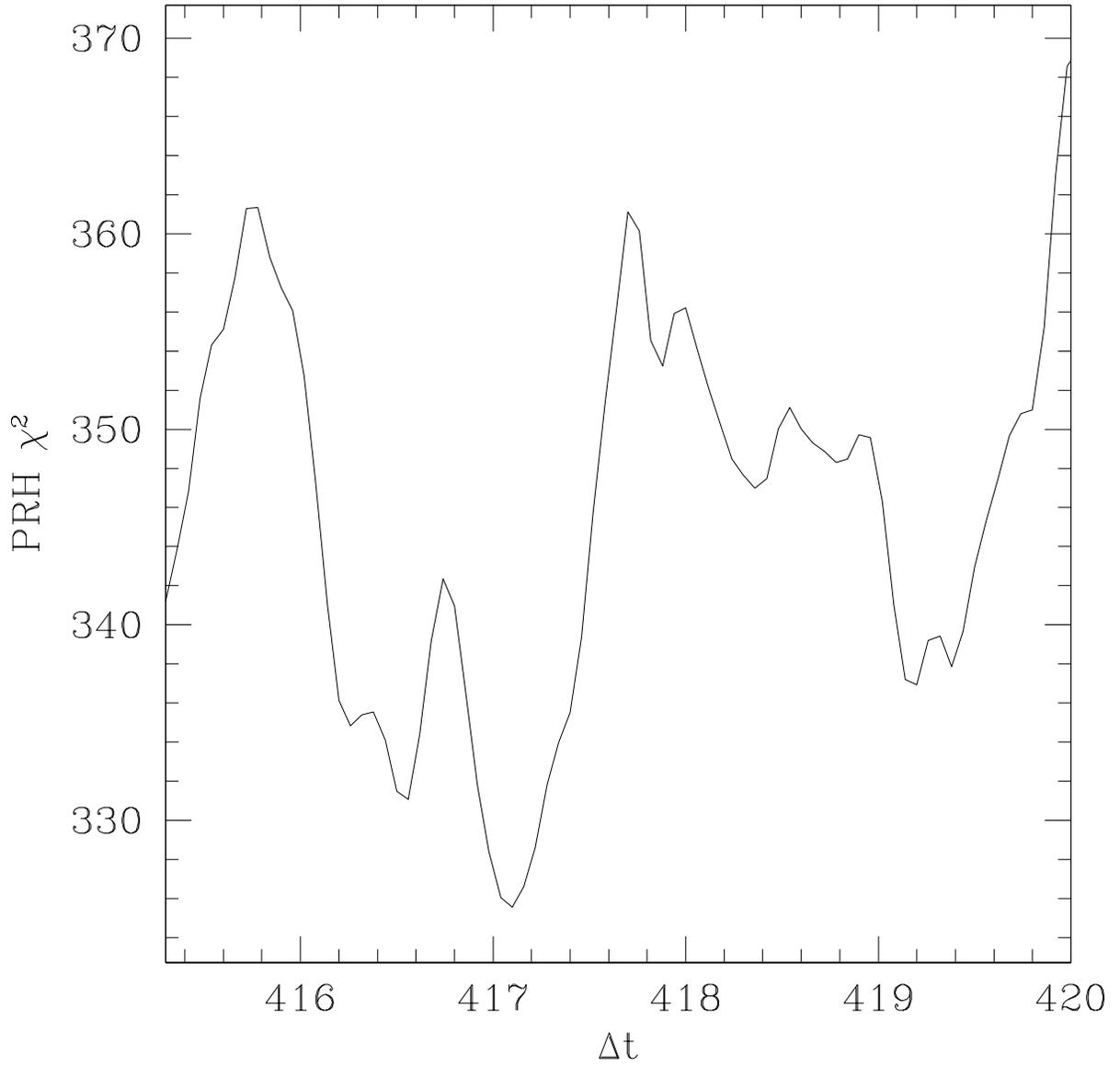}
\caption{The PRH $\chi^2$ calculated as a function of lag for the QuOC1 and
QuOC2 data. According to theory, the minimum of this curve occurs for the
most probable lag, and thus is a determination of the time delay. Notice
that the curve is reasonably symmetrical around the 417.1 day minimum.}
\label{fig2}
\end{figure}

\begin{figure}[t]
\plotone{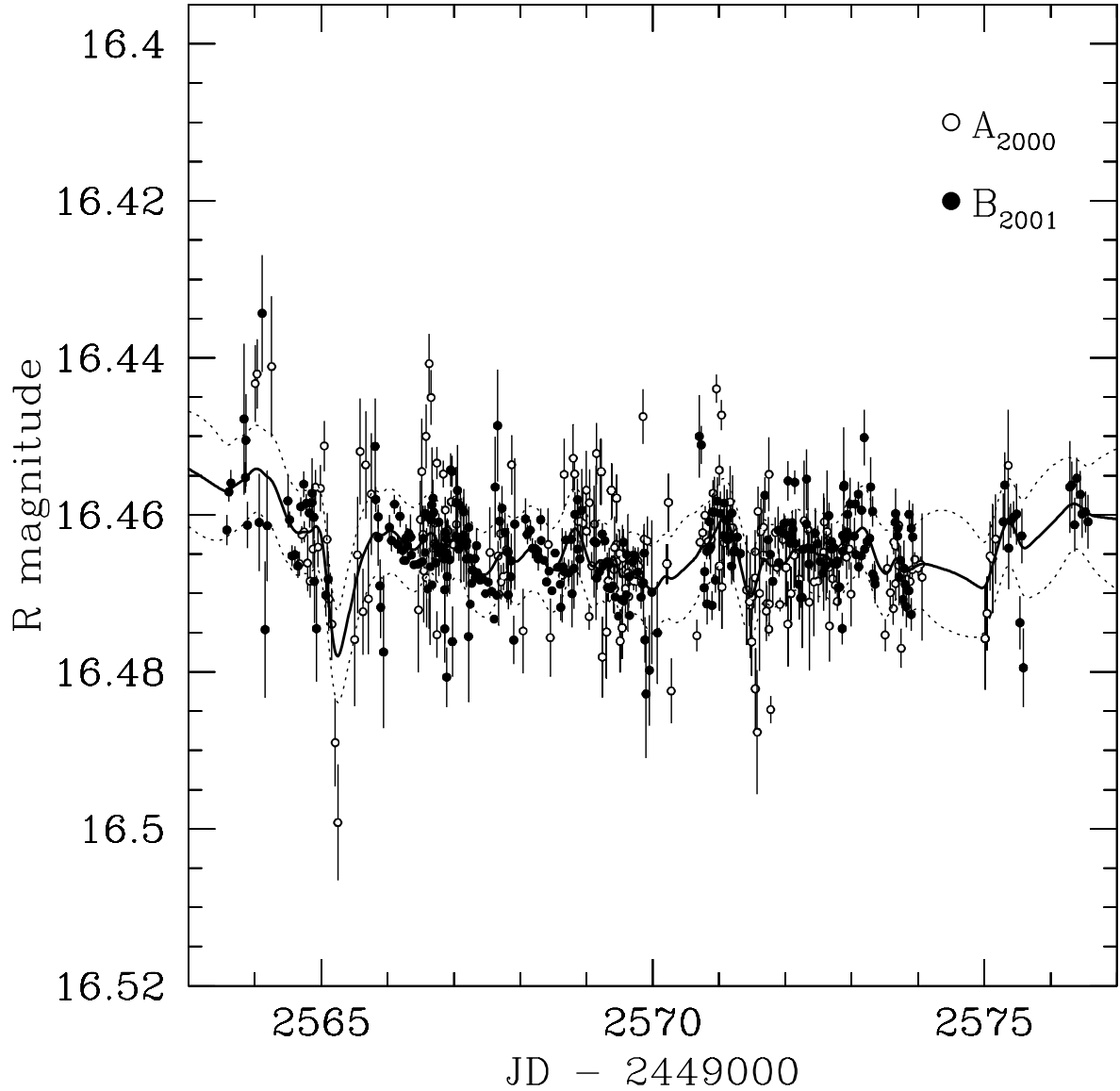}
\caption{The comparison of brightness fluctuations measured for image A
during Jan 2000 (open circles) with fluctuations measured for image B in 
March 2001 (filled circles), for a 417.1 day lag. Any significant
differences would be evidence for microlensing. The error ``snake''
corresponds to expected $1 \sigma$ departures from a mean trend, according to
Gaussian statistics.}
\label{fig3}
\end{figure}

\begin{figure}[t]
\plotone{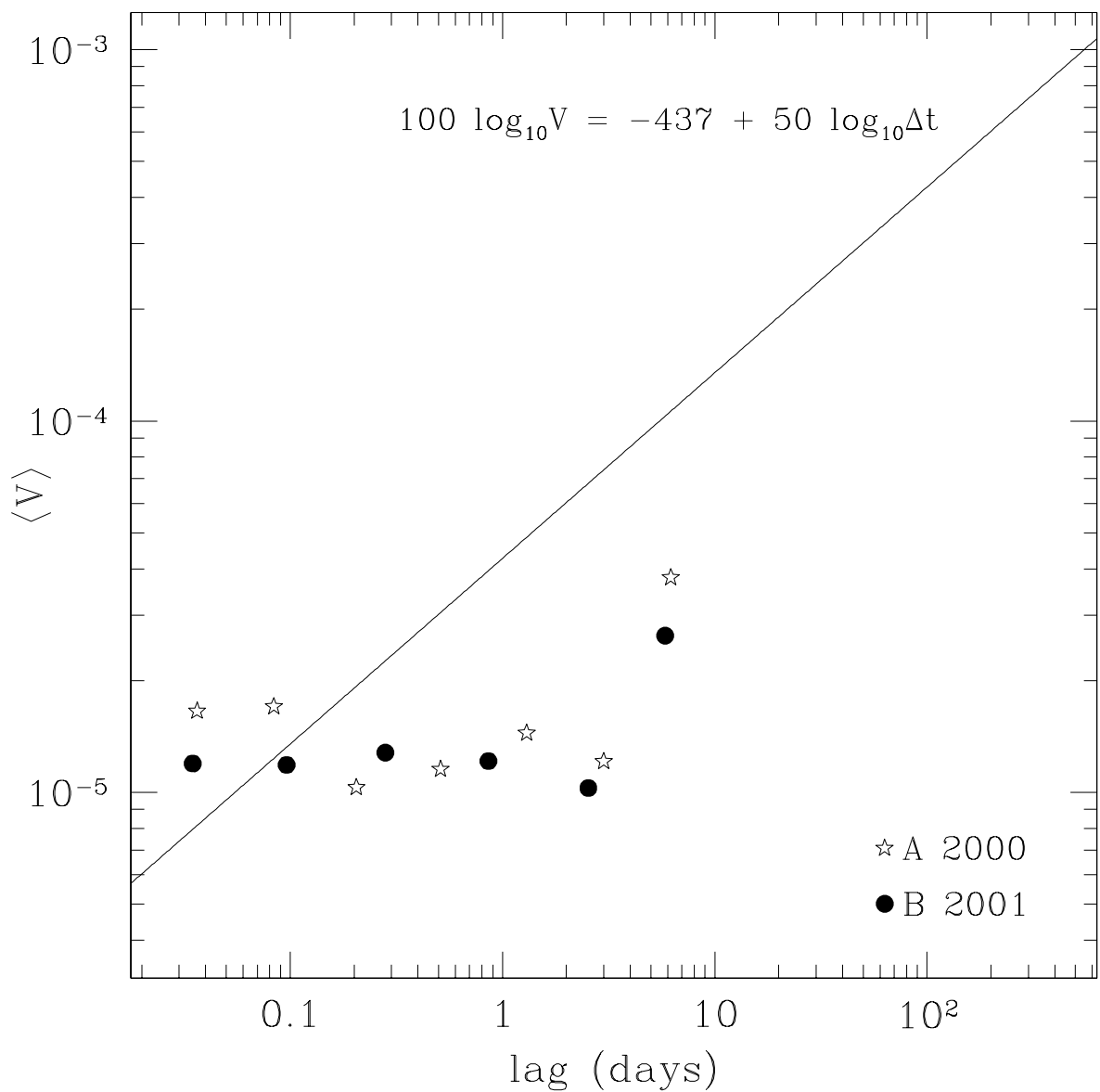}
\caption{The structure function for the observed quasar brightness
fluctuations. Filled circles show the structure function for image A
brightness in Jan 2000, and open stars are for B in March 2001.
The solid curve shows the structure function measured by Colley and Schild
(2000) for 1995 data. Because the circles and stars seem to agree so well,
we presume that plotted points show that the quasar brightness fluctuations
estimated in the two seasons agree, but the trend of the new data are
significantly different from the 1995 data. This indicates that the quasar
brightness fluctuations are statistically non-stationary. Note that the new
point for lags near 1 year are far off scale and are not plotted, again a
product of the non-stationary statistics.}
\label{fig4}
\end{figure}

\begin{figure}[t]
\plotone{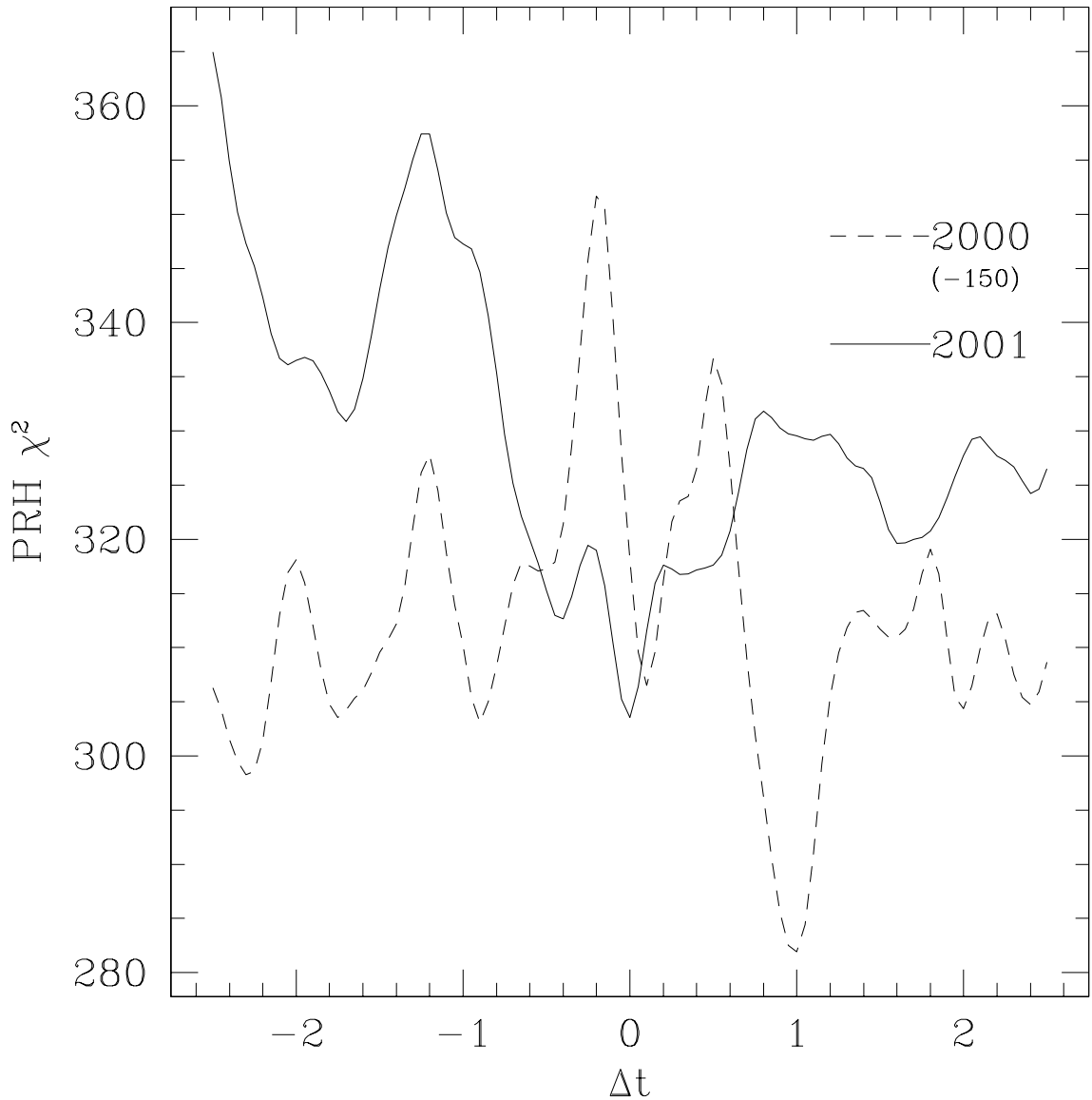}
\caption{The cross-correlation for zero lag. Careful inspection of Fig.~1
shows a surprising agreement of the many small peaks and valleys of the
objectively measured mean brightness records at zero lag. A similar but
less pronounced effect is seen for the QuOC1 data for Jan 2000. Thus we
plot the PRH method calculation separately for the Jan 2000 data (dashed
curve) and for the March 2001 data (solid curve). Note that 2000 data are
fewer in number and the curve has been shifted upward to fit on the
$\chi^2$ scale for 2001. A minimum with a FWHM of 4 hours is found for 
data in both seasons, but local minima are also found for $\pm$ 1 and $\pm$
2 days.}
\label{fig5}
\end{figure}

\end{document}